\documentclass[journal=jpccck,manuscript=article]{achemso}
\usepackage[version=3]{mhchem} % Formula subscripts using \ce{}
\usepackage{subfigure}
\usepackage{color}

\author{S. Appalakondaiah}
\affiliation{Advanced Centre of Research in High Energy Materials (ACRHEM),
University of Hyderabad, Prof. C. R. Rao Road, Gachibowli, Hyderabad - 500 046, India.}
\author{G. Vaitheeswaran}
\email{gvaithee@gmail.com}
\phone{+91-40-23138709}
\fax{+91-40-23012800}
\affiliation{Advanced Centre of Research in High Energy Materials (ACRHEM),
University of Hyderabad, Prof. C. R. Rao Road, Gachibowli, Hyderabad - 500 046, India.}
\author{S. Leb\`egue}
\affiliation {Laboratoire de Cristallographie, R\'esonance Magn\'etique et Mod\'elisations (CRM2, UMR CNRS 7036), Institut Jean Barriol, Universit\'e de Lorraine, BP 239, Boulevard des Aiguillettes, 54506 Vandoeuvre-l\`es-Nancy, France. }

\title{Dispersion Corrected Structural Properties and Quasiparticle Band Gaps of Several Organic Energetic Solids}

\begin{document}

\begin{abstract}
We have performed {\it ab initio} calculations for a series of energetic solids to explore their structural and electronic properties.
To evaluate the ground state volume of these molecular solids,  different dispersion correction methods were accounted in DFT,  namely the Tkatchenko-Scheffler method (with and without self-consistent screening),
Grimme's methods (D2, D3(BJ)) and the vdW-DF method. Our results reveals that  dispersion correction methods are essential in understanding these complex structures with van der Waals interactions and hydrogen bonding.
The calculated ground state volumes and bulk moduli show that the performance of each method is not unique, and therefore a careful examination is mandatory for interpreting theoretical predictions.
This work also emphasizes the importance of quasiparticle calculations in predicting the band gap, which is obtained here with the GW approximation.
 We find that the obtained band gaps are ranging from 4 to 7 eV for the different compounds, indicating their insulating nature.
 In addition, we show the essential role of quasiparticle band structure calculations to correlate the gap with the energetic properties.
\end{abstract}
\maketitle

\section{Introduction}
Energetic materials are good candidates for many civil, industrial and military applications due to their capability of rapid chemical decomposition with a large
 energy release to their surroundings by an external stimuli. Recently, a large community of researchers has shown interest on these compounds,  conducting their research with new strategies
 to improve the performance, effectively dealing with safety and develop new compounds which are less hazardous to the environment.  Currently, experiments on these
 materials are increasing in order to explore important properties such as structural, optical and chemical decomposition mechanisms. \cite{Tarzhanov, Bourne, Aluker,Aluker1, Aluker2,Tsyshevsky,Sharma1,Sharma2,Im,Yu1,Yu2,ZA,Greenfield,NO2,Mullen,Yu}
Indeed complex crystal structure and high sensitivity to a small external stimuli of these compounds present difficulties in experiments to examine the desired properties such
 as analysis of the bonding, stability, and studying the polymorphic phases of these materials.
 On the other hand, theoretical assessment on these materials is an alternative method to provide a description of the structural and electronic properties at various conditions.

Density functional theory (DFT) is an efficient computational tool, and the use of this method resulted in significant advances for various materials. Conventional DFT methods
 with existing standard approximations to the exchange-correlational functionals (modeled using the local density approximation (LDA)\cite{PPerdew} or the generalized gradient approximation (GGA)
 with different flavors such as PBE\cite{Perdew} or PW91\cite{Wang}) are in excellent agreement with experiments for covalent and metallic systems. However standard DFT
  has two major general drawbacks in: 1) treating  long range dispersive interactions  and 2) predicting (and reproducing) the band gap from Kohn-Sham eigenvalues.
  In fact, these two problems play an important role in energetic materials to determine the structure, bonding, and sensitivities, as described below.

 \begin{figure*}
\centering
{\includegraphics[width=6in,clip]{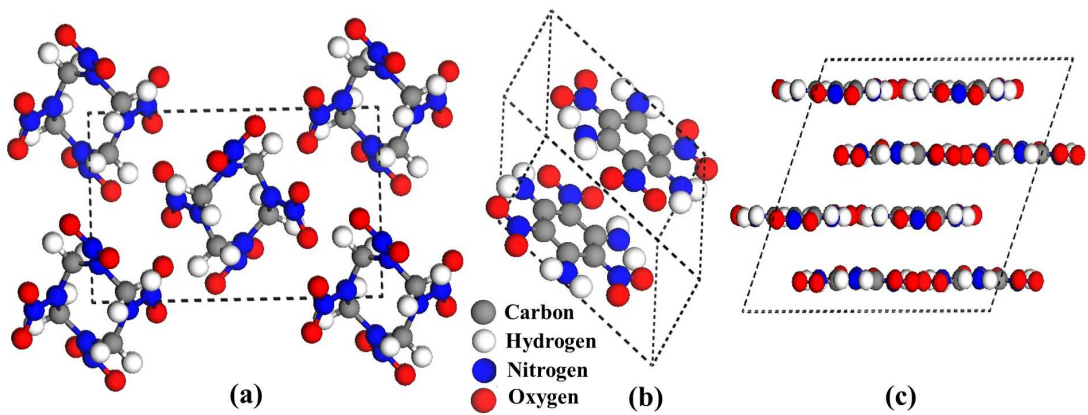}}
\caption{(Color online) Experimental crystal structures: (a) $\beta$-HMX, (b) unit cell of TATB  and (c) super cell of TATB. }
\end{figure*}

\par It is known that most of the energetic materials are molecular solids (including layered compounds), where the crystal packing between adjacent molecules
 is governed by both van der Waals interactions and hydrogen bonding (as particulary observed in C-H-N-O based molecular crystals).  For example, the structures of two widely used high energetic
  materials, $\beta$-HMX and TATB, are shown in Figure .1.
  Earlier, Byrd and co-workers\cite{Byrd1,Byrd2} studied different energetic materials with conventional DFT methods using various exchange-correlational functionals (such as LDA, PBE, and PW91).
    Their results clearly show large errors in reproducing the ground state volume. The discrepancy between theory and experiments leads to
     difference in the density ($\rho$$_0$), which has drastic effects in reproducing the energetic properties of these solids in terms of detonation velocity
      (D$_v$ $\propto$ $\rho_0$) and pressure (D$_p$$\propto$ $\rho_0$$^2$). This deviation in reproducing ground state volumes  between conventional DFT methods and experiments
       was expected due to the fact that non local dispersive interactions were not accounted in their calculations.
To correct this discrepancy between available experiments and DFT results for the ground state volumes, it is essential to include dispersive interactions in the calculations.
There have been promising attempts made to fix this shortcoming, and  considerable benchmarking studies are available with advanced methods for noble gas solids, layered, sparse
 and molecular crystals.\cite{v1,v2,v3,v4,v5,v6,v7,v8,v9,v10,v11} In the case of energetic materials, Sorescu et al\cite{Sorescu} performed calculations using Grimme's\cite{Grimme} method
 for 10 systems and obtained improved results with an error for the ground state volumes ranging from approximately -3 to 4\%.
 In addition, Landervilli et al. reported ground state volumes and bulk moduli for some energetic materials using  Neuman and Perrin approach.\cite{Lander} From our previous studies on
 nitromethane and FOX-7 solids,\cite{SA1,SA2} we also found good agreement using Grimme and TS methods.
 However, it is observed that although the mentioned dispersion corrected methods overcomes most of
 the difficulty in reproducing ground state volumes, the relative errors still show a mean absolute deviation of about 3 \%.

 \begin{table*}
\caption{Summary of simulated structures used in the present work. All the calculations performed at kinetic energy
cut off of 800 eV. Here, CS: Crystal structure, SG: space group, Z: no of formulas  per unit cell, N: no of atoms per
unit cell, and V : volume in \AA$^3$ }
%\begin{ruledtabular}
\begin{tabular}{cccccc} \hline
Compound &  CS  & SG &  Z (N) & V \\ \hline
$\beta$-HMX (C$_4$H$_8$N$_8$O$_8$)$^a$& Monoclinic & P2$_1$/c & 4 (56) &  519.39\\
TATB (C$_6$H$_6$N$_6$O$_6$)$^b$& Triclinic & P\={1} & 2 (48) & 442.49\\
NTO (C$_2$H$_2$N$_4$O$_3$)$^c$& Monoclinic & P2$_1$/c & 4 (44) & 450.29\\
TEX (C$_6$H$_6$N$_4$O$_8$)$^d$& Triclinic & P\={1} & 2 (48) & 433.64\\
TAG-MNT (C$_3$H$_{12}$N$_{12}$O$_2$)$^e$& Triclinic & P\={1} & 2 (58) & 525.60\\
FOX-7 (C$_2$H$_4$N$_4$O$_4$)$^f$& Monoclinic & P2$_1$/c & 4 (56) & 515.89 \\
%NM (CH$_3$NO$_2$)& Orthorhombic & P2$_1$2$_1$2$_1$ & 4 (28) & 29
\hline
\end{tabular} \\
$a:$ Ref. \citenum{1}, $b:$ Ref. \citenum{2}, $c:$ Ref. \citenum{3}, $d:$ Ref. \citenum{4},$e:$ Ref. \citenum{5}, $f:$ Ref. \citenum{6} . \label{CS}
\end{table*}

Another important property concerning energetic materials is the electronic band gap.
Earlier, Kuklja and coworkers estimated the narrowing of the band gap using molecular level calculations for different energetic materials
from ambient to high pressure and studied the key role of band gap as well as electronic excitations on the decomposition process.\cite{Kuklja1,Kuklja2,Kuklja3,Kuklja4,Kuklja5}
 A few other studies were also proposed to model the link between the impact sensitivity and the electronic band gap with the PBE functional
 (for polymorphs of HMX, CL20 and the nitraromatic series MATB, DATB, and TATB) and reported that the lower is the band gap, then higher
 is the sensitivity.\cite{Zhu,Hong} But the well known disagreement of the computed band gap using standard DFT functionals leads to an inaccurate estimation of the properties related to the band gap either in molecular or solid systems.
 However from our previous studies on some energetic materials (solid nitromethane, FOX-7, cynauric triazide)\cite{SA1,SA2,SA3}
  and another theoretical study on TATB  by Fedorov et al\cite{Igor}, it is known that the GW approximation\cite{gap1,gap2,gap3,gap4,gap5,gap6} can overcome this problem.

  Therefore in the present work, we aim to carry out theoretical study of C-H-N-O based energetic solids for understanding the relative errors in structural properties using various dispersion correction methods and quasiparticle band structure calculations for obtain the accurate band gaps values.  For this, we have chosen different energetic solids (namely $\beta$-HMX, TATB, NTO, TEX, TAG-MNT and FOX-7) to compare the obtained results with available experiments as well previous reports.  Here, the computed energetic solids are popularly known compounds in which the inter (intra) molecular interactions resembles with mixed nature of both van der Waals and hydrogen bonding. In addition, the quasiparticle band gaps of these compounds are not yet reported to these compounds except for TATB\cite{Igor}. The experimental crystal structure details and the corresponding information are presented in \ref{CS}. It is also to be noted that performing both structural and quasiparticle band structure calculations for the above mentioned compounds are computationally expensive (see Table 1 for structural details\cite{1,2,3,4,5,6} in the view of number of atoms/cell), and hence we confined our self with six compounds. The rest of paper is organized as follows: in section II, we describe
  the computational techniques that we used, while the results and discussions are presented in section III. Finally, a brief conclusion is given in section IV.
  
\section{Computational details}
The calculations in the present work were carried out with the projector augmented wave (PAW)\cite{paw1} implementation of VASP.\cite{vasp}
The generalized gradient approximation (GGA) in the PBE parameterization was considered as the exchange correlational functional\cite{Perdew}.
Structural optimizations were achieved by setting the convergence criteria for total energies are below 5e-06 eV, residual forces to be less than 1e-3 eV/$\AA$
 and stresses are limited to 0.02 GPa. To correct the missing dispersion interactions, we have used several recently proposed methods such as PBE functional
 with pair potential methods D2, D3 (BJ), TS , TS+SCS and density functional (vdW-DF) methods as implemented in the VASP code. The details of the implementations  and usage of these  methods can be found elsewhere(Ref. \citenum{Grimme,D3,TS,TS+SCS,DF}). To obtain accurate band gaps, we have used the GW approximation\cite{GW2,GW3}. For getting accurate quasiparticle eigenvalues, we used 200 bands for the summation over
 the bands in the polarizability and the self-energy formulas, and the polarizability matrices were calculated up to a cut-off of 200 eV.
\section{Results and discussions}
\subsection{Ground state volume and bulk modulus}
\begin{figure*}
\subfigure[]{\includegraphics[width=3.2in,clip]{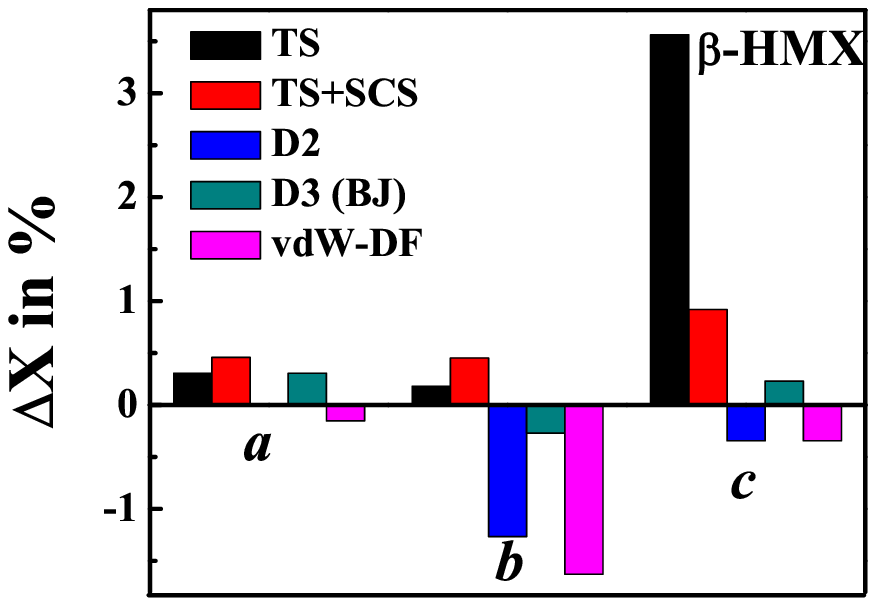}}
\subfigure[]{\includegraphics[width=3.2in,clip]{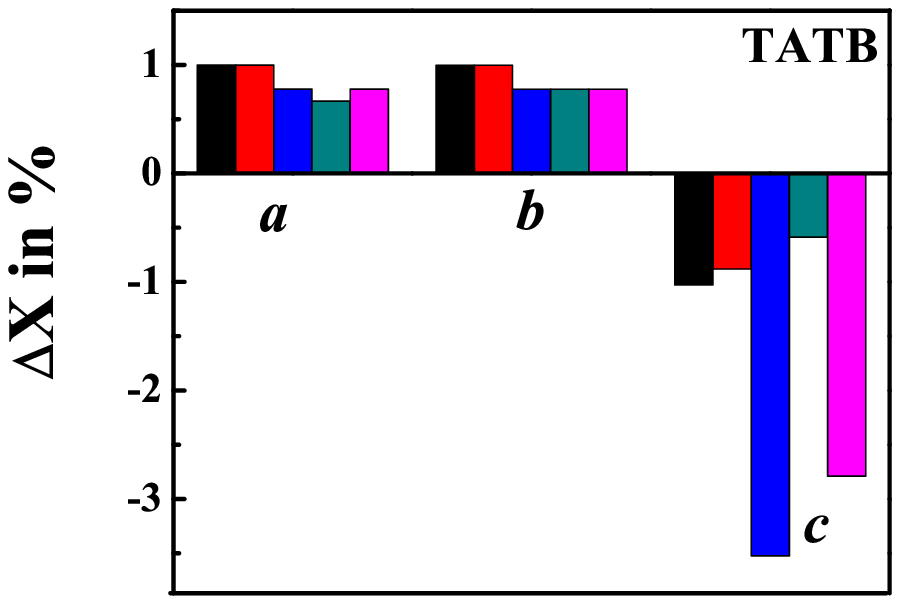}}
\subfigure[]{\includegraphics[width=3.2in,clip]{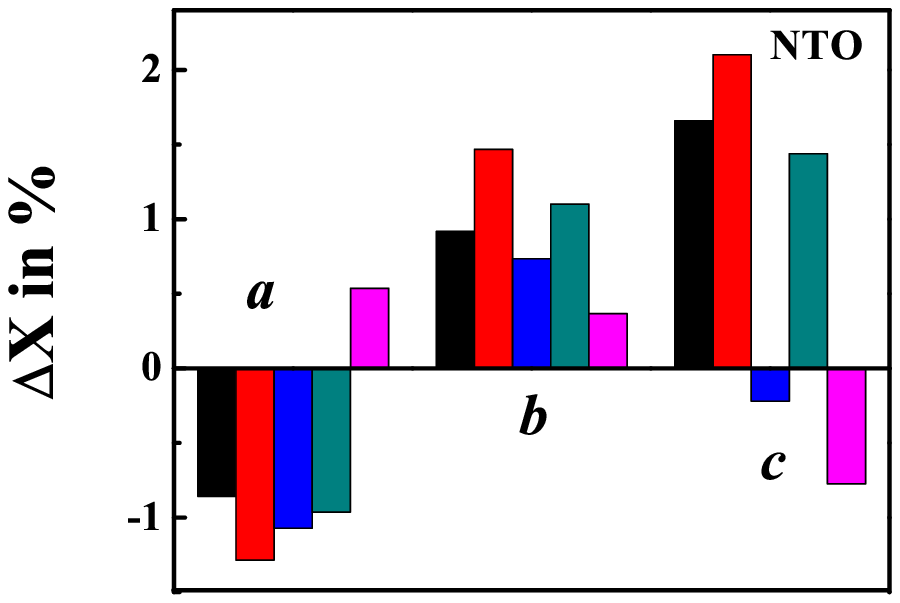}}
\subfigure[]{\includegraphics[width=3.2in,clip]{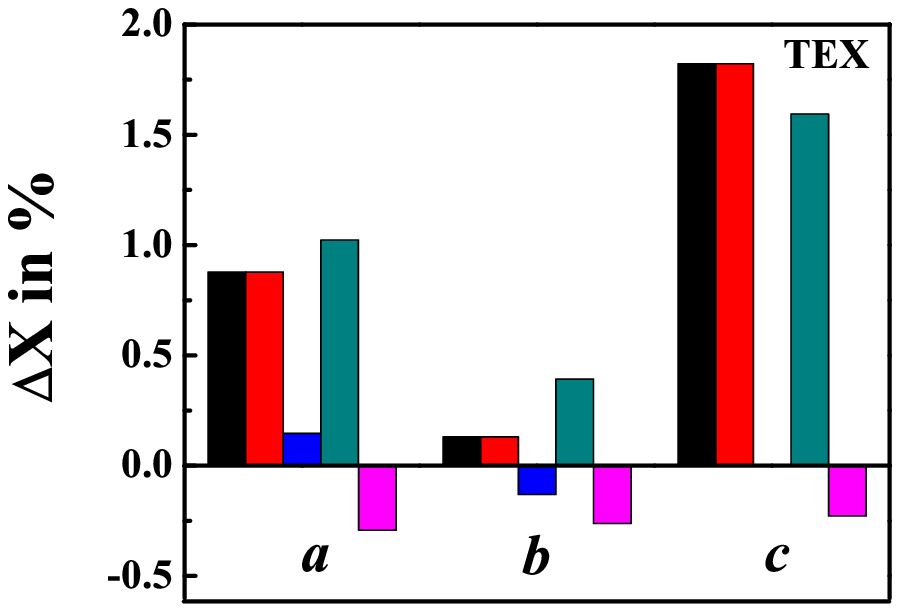}}
\subfigure[]{\includegraphics[width=3.2in,clip]{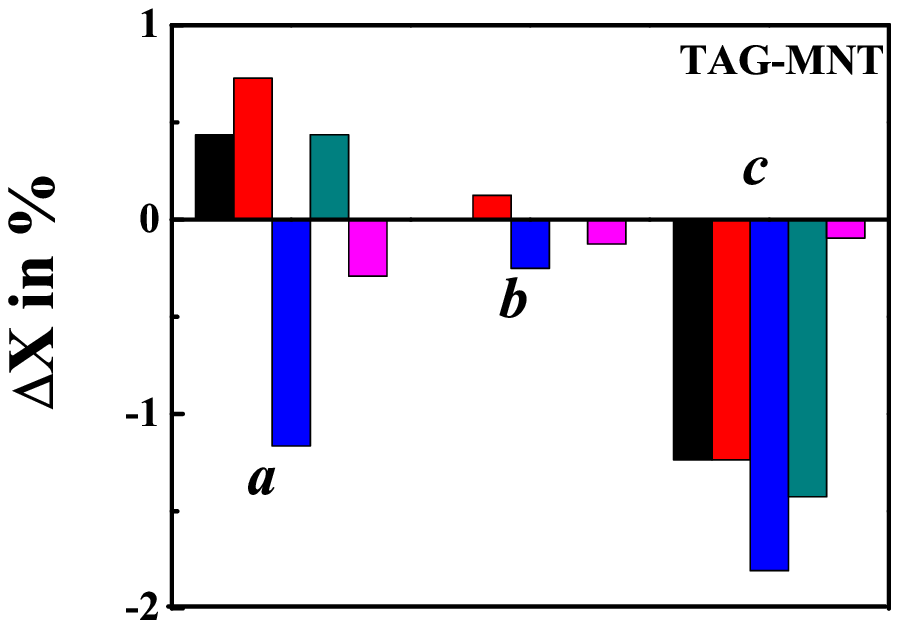}}
\subfigure[]{\includegraphics[width=3.2in,clip]{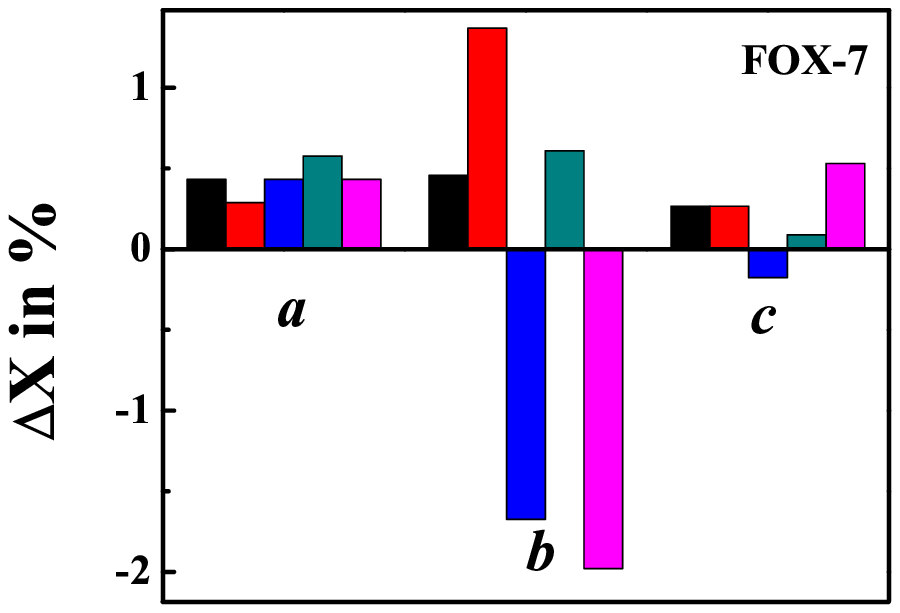}}
\caption{(Color online) Calculated relative errors (in \%) of lattice parameters ($a$, $b$, and $c$) for organic energetic solids. (a). $\beta$-HMX, (b). TATB, (c). NTO, (d). TEX, (e). TAG-MNT and (f). FOX-7.
}\label{abc}
\end{figure*}

Firstly, geometry optimizations were performed for six organic energetic solids with the various methods mentioned in the previous section. The complete list of resulting lattice parameters, angles,
 and volumes along with experimental results for all the compounds are presented in Supporting Information. As expected, the PBE functional overestimated the lattice parameters (around 5 \%) and
 in some cases the monoclinic and triclinic angles are slightly too low (around 1 \%). Overall the PBE functional overestimated the volumes ranging from 7 to 15 \% as compared with experiments
 for the tested compounds(see Supporting Information Table. 1). On the other hand, various dispersion corrections to the PBE functional lead to an improvement in the structural parameters and the obtained results are in good agreement
 with experiments. Now, we will take a look into the accurate dispersion correction method for organic energetic solids in the present work. In order to enable this, we tested the percentage of relative
 error ($\triangle$X =((X$_{cal}$-X$_{exp}$)/X$_{exp}$)$\times$100) in the lattice parameters, volumes and densities using various dispersion corrections. The obtained $\triangle$X with dispersion
 corrections for lattice parameters are shown in \ref{abc}(a)-(f) and $\triangle$X for volumes are illustrated in \ref{vol}.
From \ref{abc}, it is clear that  $\triangle$X (X= $a$, $b$, and $c$ ) for lattice parameters using different approaches varies from -3 to +3\%  for all compounds, and none of the individual method achieved consistently minimal error results for the tested systems. Obviously this may be expected due to the mixed nature of vdW forces and hydrogen bonding in these organic energetic solids. For example, the considered compounds TATB as well as FOX-7 are layered solids and layers are dominated with vdW forces (along $c$-axis for TATB and $b$-axis for FOX-7) and hydrogen bonding is also an inherent accessible (C/N...O-H bonding pairs) for stability of these structures.

To further understand the performance of each functional, we considered the equilibrium volumes obtained with the various methods. Turning now to \ref{vol}, the mean absolute deviation (MAD) of all dispersion correction methods is almost within a 1-2 \% deviation with experiments and $\triangle$V in the equilibrium volumes are scattered with each method used. For instance, by considering a case of less than 1\% relative error at equilibrium volume, we found a good agreement with: the TS and D3(BJ) methods for $\beta$-HMX; the TS, TS+SCS and D3(BJ)
 methods for TATB; the D2 and vdW-DF methods for NTO and TEX systems. In contrast to these, all the dispersion corrected schemes (except D2) were found to show $<$ 1\% deviation for TAG-MNT and
 above 1\% of deviation for FOX-7 in comparison with experiments.
 In addition to the present work,  Sorescu et al\cite{Sorescu} found -1.18 \% for $\beta$-HMX, -3.24 \% for TATB, and -0.48 \% for FOX-7 using the PBE-D2 method;
a deviation of -2 \% was observed for $\beta$-HMX and TATB using the Neuman and Perrin approach \cite{Lander};  Wu et al\cite{Qiong} reported the ground state
 volume of NTO with a deviation of -0.6 \% using the D2 method. At this point, these observations indicates that dispersion correction methods replicate the volume with experiments, but the performance of each method is not unique for different compounds.

\begin{table*}
\caption{The calculated ground state volumes V (in \AA),  \% of relative error ($\triangle$V) with experiments (see \ref{CS} for experimental volumes)  with standard DFT method (PBE) and  different dispersion corrected methods.}
%\begin{ruledtabular}
\begin{tabular}{ccccccccccc} \hline
      &  & PBE  &      +TS &    +TS-SCS  &   +D2 &    +D3(BJ)  &+vdW-DF&    \\ \hline
$\beta$-HMX&V & 578.70  & 524.05  & 528.38  & 510.75  & 520.56  & 507.26 \\
&$\triangle$V & +11.42& +0.90 & +1.73 & -1.66 & +0.22 & -2.33  \\

 TATB  &V  & 534.28  & 443.92  & 445.91  & 430.32  & 445.47  & 433.40   \\
&$\triangle$V& +20.74& +0.32 & +0.77 & -2.75 & +0.67 & -2.05  \\

 NTO    & V  & 504.87  & 458.81  & 462.09  & 448.86  & 458.99  & 450.68   \\
 &$\triangle$V & +12.12 & +1.89 & +2.62 & -0.32 & +1.93 & +0.09  \\

 TEX   &V      & 501.40  & 444.86  & 447.57  & 433.16  & 446.86  & 430.23   \\
&$\triangle$V& +15.63 & +2.59 & +3.21 & -0.11 & +3.04 & -0.77  \\

 TAG-MNT   & V&     565.95  & 523.03  & 525.55  & 508.36  & 521.41  & 522.93   \\
&$\triangle$V & +07.6& -0.49 & $\sim$0.00 & -3.28 & -0.80  & -0.51 \\

 FOX-7  & V   & 598.61  & 522.54  & 526.78  & 508.62  & 521.94   & 510.19 \\
&$\triangle$V   & +16.03& +1.28 & +2.11 & -1.41 &  +1.10 &  +1.17\\ \hline
 MAD  &  & 14.23  & 1.24 & 1.74 & 1.59 & 1.30 & 1.14  \\ \hline

% NM   & V        & 322.84  & 285.95 &289.90  & 276.35 & 273.31  & 281.46 &286.37 & 275.31 \\
 %     & e        & +17.26  & +3.86 & +5.30 &  +0.73  & -0.72 & +2.24 \\

   \end{tabular}
%\end{ruledtabular}
\label{vol}
\end{table*}

\begin{figure*}
{\includegraphics[width=3.2in,clip]{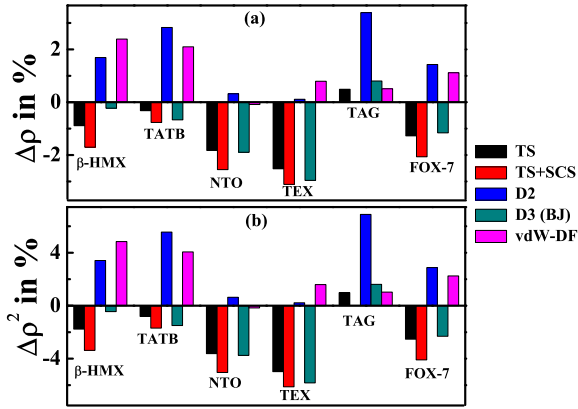}}

\caption{(Color online) Calculated relative errors (in \%) of density ($\rho$) for organic energetic solids. Here (a)$\triangle \rho$ and (b)$\triangle \rho^2$ are tested with various dispersion correction methods in the present work.}\label{Density}
\end{figure*}

 We next move to analyse the calculated density, which is an intrinsic parameter to determine explosive properties in energetic solids. As mentioned in the introduction D$_v$ $\propto$ $\rho$ and D$_p$$\propto$ $\rho$$^2$ for energetic solids; in \ref{Density}(a),(b), we plot the obtained $\triangle \rho$ as well $\triangle \rho^2$ for all compounds. Similar to ground state volume, $\triangle \rho$ also varies around -2 to 2 \% with various methods and $\triangle \rho^2$ was enhanced to  -5 to 5 \% with various dispersion correction methods. However, this tendency of relative error in $\rho^2$ should be considered towards predicting explosive properties such as detonation pressure.

.

\begin{table*}
\caption{The calculated Bulk moduli and its pressure derivative of energetic materials with standard DFT method (PBE) and  different dispersion corrected methods.}
%\begin{ruledtabular}
\begin{tabular}{ccccccccccc} \hline
Compound    & Method   & B$_{0}$ (B') & Exp \\ \hline
$\beta$-HMX & PBE   & 9.24 (7.27)   \\
&              TS   & 16.07 (7.84) \\
&          TS+SCS   & 14.03 (8.06) & 12.4 (10.4)\cite{Choong}\\ % at 528.04 $\AA^3$ \\
&              D2   & 17.34 (7.85) & 21.0 (7.4)\cite{Gump} \\ %at 515.40 $\AA^3$ \\
&              D3 (BJ)   & 15.93 (7.38) \\
&          vdW-DF  & 20.98 (8.58) \\ \\

TATB        & PBE   & 3.49 (12.66) \\
            & TS    & 21.12 (6.30) \\
         &TS+SCS    & 17.78 (7.74) & 15.7 (8.0)\cite{Lewis} \\ % at 442.5 $\AA^3$ \\
            & D2    & 14.12 (10.82) \\
            & D3 (BJ)    & 14.27 (7.91) \\
         &vdW-DF   & 20.95 (7.85)   \\ \\

TAG-MNT         & PBE  & 5.24 (11.42) \\
            & TS   & 20.37 (3.47) \\
         &TS+SCS   & 20.65 (3.37) & 14.6 (4.8)\cite{McWilliams} \\ % at 533.00 $\AA^3$ \\
            &D2    & 18.72 (4.30) \\
            &D3 (BJ)   & 19.59 (3.35) \\
         &vdW-DF  & 21.00 (3.83)  \\ \\

NTO        & PBE  & 11.32 (6.65) \\
            & TS   & 21.60 (5.27) \\
         &TS+SCS   & 20.29 (5.81)  \\
            &D2    & 18.06 (6.20) \\
            &D3 (BJ)    & 18.68 (6.15) \\
         &vdW-DF  & 25.15 (4.26) \\ \\

TEX        & PBE  &  7.26 (8.41) \\
            & TS   & 15.70 (7.30) \\
         &TS+SCS   & 14.38 (6.89)  \\
            &D2    & 16.91 (7.30) \\
            &D3 (BJ)   & 13.51 (8.14) \\
         &vdW-DF  & 20.66 (7.75) \\ \hline

  \end{tabular}
%\end{ruledtabular}
\label{BM}
\end{table*}

  As a next step we also calculated the bulk moduli of organic energetic solids to check the performance of various dispersion correction methods. Besides the ground state volume and density, a similar situation was observed even in the computed bulk moduli for these compounds using various methods.  \ref{BM} shows the obtained bulk moduli (B$_0$) and their
 pressure derivatives (B') along with available experimental values. Here, B$_0$ and B' were computed with the Murnaghan equation of state, where the P-V data were fitted
 from 0 to 5 GPa with a step size of 0.5 GPa. By comparing the computed B$_0$  from various dispersion correction DFT methods with the available experimental values, one can understand that the obtained values fall in the range of experiments.
  We also note that the present computed ground state volume, bulk moduli and pressure derivatives for $\beta$ - HMX and TATB
 are comparable with previous reports using D2 as well as Neuman and Perrin approaches,\cite{Sorescu,Lander} and to the best of our knowledge experimental or any dispersion corrected DFT calculations
 for NTO, TEX compounds are not available.

  From the corresponding results of ground state volumes, densities and bulk moduli with different dispersion schemes,  it is
  observed that the \textit {best performing} dispersion method for C-H-N-O based energetic solids particularly depends on the chemical
  environment of the respective crystal. In other words, the chemical bonding (intra- as well as inter molecular interactions) in these C-H-N-O based energetic solids are significantly dominated with mixed of van der Waals forces and hydrogen bonding, while the geometrical arrangement in these complexes are different from one another. This might be the cause that the computed relative errors are quite different for tested systems. Note that, our results are in line with recent report on high nitrogen-content energetic salts using various dispersion correction methods.\cite{Sor1} We also consider that, it would be essential to perform theoretical calculations by verifying  proper dispersion correction method to qualitative predict important explosive properties such as D$_p$ or D$_v$, and  other properties like elastic or dynamics, polymorphism.

\subsection{Quasiparticle band gaps}
\begin{figure*}
\centering
{\includegraphics[width=4in,clip]{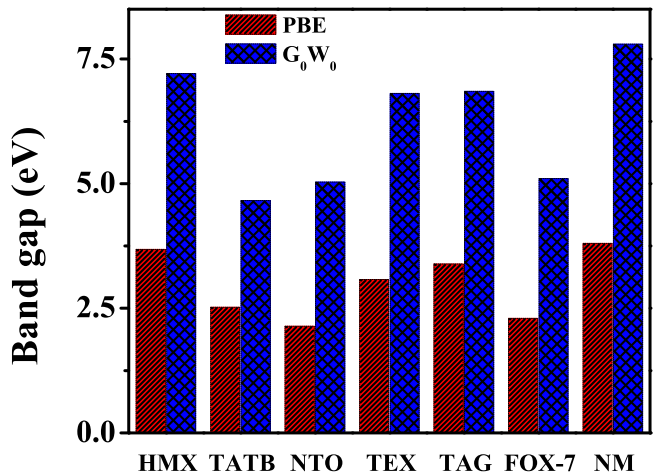}}
\caption{(Color online) Calculated G$_0$W$_0$(PBE) band gaps (in eV) for a series of energetic materials. Here the band gaps (in eV) are: for HMX: 7.21 (3.68),
TATB: 4.66 (2.52), NTO: 5.03 (2.14), TEX: 6.81 (3.08), TAG : 6.85 (3.39), FOX-7\cite{SA1}: 5.1 (2.3) and NM\cite{SA2}: 7.8 (3.8)}\label{GW}
\end{figure*}

As mentioned in Introduction, the previous models with standard GGA/LDA methods projected the key role of electronic band gap to understand the sensitivity or decomposition mechanism of energetic materials at various pressure conditions. However, it is a known fact that standard DFT methods have shortcoming in predicting/reproducing band gap due to the derivative discontinuity of the functionals and there is a  prerequisite for accurate band gap values to confirm the previously reported theoretical models of energetic solids.\cite{Kuklja1,Kuklja2,Kuklja3,Kuklja4,Kuklja5,Zhu,Hong} Therefore, to investigate the fundamental electronic band gap of these energetic solids, we have performed quasiparticle band structure calculations with the GW approximation (G$_0$W$_0$) method for the experimental crystal structures.  The computed GGA-PBE and G$_0$W$_0$ band structures are presented in Supporting Information. From the computed band structures, it is observed that bands near the Fermi level are nearly flat with both methods, and as compared with PBE band structures, a wide upshift in the conduction band minimum was noticed after taking the quasiparticle correction with PBE functional. Specifically, we found nearly 40 \% increment in the computed band gaps for all compounds which substantiate the necessity of quasiparticle corrections to calculate the electronic band gap of these systems. A systematic comparison of our calculated band gaps with G$_0$W$_0$ as well as PBE is presented in \ref{GW}. From this, we noticed that all these C-H-N-O based energetic materials are wide band gap insulators, ranging from 4 to 8 eV (310 to 155 nm), while PBE results yield only from 2 to 4 eV (600 to 300 nm). Since, very limited information is available regarding experimental band gaps (only available for $\beta$-HMX and TATB in the present work) of energetic solids, we directly compare our quasiparticle band gaps with available values from absorption spectra.  For, $\beta$- HMX, the fundamental optical absorption peaks exhibited near to UV region from experiments,\cite{UV,smit} ranging from  at 5.32-6.39 eV (233-194), while the obtained quasiparticle is 7.1 eV (172 nm). The computed G$_0$W$_0$ band gap of TATB (4.66 eV) is in excellent agreement with previously reported value of 4.29 eV\cite{Igor} and relatively low when compare to the experimental value (6.5 eV) using X-ray absorption spectroscopy \cite{Kakar}. From these it is found that, the obtained G$_0$W$_0$ band gap of $\beta$-HMX is overestimated (value of 1.69-0.62 eV) with experiments, whereas TATB value is lower than experiments by 1.8 eV in the present work.  To the best of our knowledge, the band gaps or absorption spectra for NTO, TEX, TAG-MNT and FOX-7 are not available. It is to be noted that photoemission experiments are necessary to examine the real band gap for any material and the present results may provide useful inputs to perform experiments. On the other hand, the differences between obtained G$_0$W$_0$ band gaps and available absorption regions of energetic solids might be the major role of exciton effects, which is beyond the scope of this paper. For instance, electronic excitations were observed experimentally for few of the energetic solids such as nitromethane (at 226 nm), RDX (at 230 nm), dimethylnitramine (at 236 nm) and PETN (at 193 nm)\cite{UV,NO2,Mullen,Yu}.

Besides the comparison of band gaps with experiments, the results from quasiparticle corrections with PBE give insight the necessity of theoretical models to go beyond GGA methods. For example, a quantitative assessment of metallization of pure/defect energetic solids may be expected at high pressures using quasiparticle methods as compared with standard PBE method (see Ref. \citenum{met1,met2,met3}). Also, considering other previous theoretical model on $\textit {band gap vs impact sensitivity}$ of energetic solids, it is necessary to revisit the same model with higher than standard GGA/LDA level.

\section{Conclusions}

In summary, we have performed first principles calculations for several energetic materials concerning their structural and electronic properties, which strongly suggest the necessity
 to go beyond standard DFT. Firstly, we have shown that dispersion corrected methods are necessary to provide an improvement in the calculated volumes to compare with experiments.
 Furthermore, the mean absolute deviation in theoretical volumes for all tested methods are similar (around 1.2 to 1.8\%) and the relative errors on structural properties calculated using each method vary by an order of $\sim(0-3.5)$ \% for different compounds in the present work. Additionally, similar situation was observed for other volume-dependent properties such as density and bulk moduli. The present observations clearly demonstrated that one has to test with all the dispersion corrected DFT methods for C-H-N-O based energetic solids before concluding the ground state properties, so that other related explosive properties (detonation pressure or velocity), polymorphic phase transitions, mechanical properties and many more can be predicted with reasonable accuracy.
  Besides this, we have calculated the dispersion corrected bulk moduli in the low pressure region which suggests the soft nature of these materials.
 Second, the use of quasiparticle band structure calculations indicate a  significant enhancement (nearly 40\%) in the band gap in comparison with the results obtained with the PBE functional,
 which emphasizes the wide band gap insulating nature of energetic solids. The calculated fundamental band gap for $\beta$-HMX as well as TATB are far from the reported experimental optical absorption values, which might indicate that excitation binding energies play an important role in these compounds. By comparing the present G$_0$W$_0$ band gaps with available previous reports on other energetic solids, it open up the necessary of future calculations including electronic excitons using G$_0$W$_0$ or higher methods to examine the decomposition mechanism of these energetic solids and possible metallization under pressure. It could also be essential to revisit  the previous model of $\emph{band gap with impact sensitivities}$ for energetic solids within similar structure or polymorphic phases beyond standard PBE method. Finally, the present work will stimulate experiments to perform photoemission study on the above compounds.

\section{Acknowledgements}
S. A. would like to thank DRDO through ACRHEM for financial support. S. A. and G. V. thank CMSD, University of Hyderabad, for providing computational facilities. S. L. acknowledge the acces to HPC resources from GENCI-CCRT/CINES (Grant x2015-085106). Part of the calculations were performed in the Computing Centre of the Slovak
Academy of Sciences using the supercomputing infrastructure acquired in
project ITMS 26230120002 and 26210120002 (Slovak infrastructure for high-performance computing)
supported by the Research \& Development Operational Programme funded by the
ERDF.
$^*$\emph{Author for Correspondence, E-mail: gvaithee@gmail.com}
{}
%\begin{tocentry}
%\includegraphics[width=2in,clip]{10}
%PBE and G$_0$W$_0$ band gaps of C-H-N-O energetic solids .
%\end{tocentry}
\end{document}